\documentclass[a4paper]{article}

\usepackage{INTERSPEECH2022}
\usepackage{graphicx,xcolor}
\usepackage{slashbox}
\usepackage{multirow}
\usepackage{amssymb}
\usepackage{bbding}
\usepackage{algorithm}
\usepackage[noend]{algorithmic}

\usepackage{float}
\usepackage[flushleft]{threeparttable}

\title{Leveraging Real Conversational Data for\\Multi-Channel Continuous Speech Separation}
\name{Xiaofei Wang$^*$\thanks{$^*$The authors equally contributed to this work. }, Dongmei Wang$^*$, Naoyuki Kanda, Sefik Emre Eskimez, Takuya Yoshioka}
\address{Microsoft Cloud+AI, USA} 
\email{\{xiaofewa, dowan, nakanda, seeskime, tayoshio\}@microsoft.com}

\begin{document}

\maketitle
\begin{abstract}
Existing multi-channel continuous speech separation (CSS) models are heavily dependent on supervised data - either simulated data which causes data mismatch between the training and real-data testing, or the real transcribed overlapping data, which is difficult to be acquired, hindering further improvements in the conversational/meeting transcription tasks.
In this paper, we propose a three-stage training scheme for the CSS model that can leverage both supervised data and extra large-scale unsupervised real-world conversational data.
The scheme consists of two conventional training approaches---pre-training using simulated data and ASR-loss-based training using transcribed data---and a novel continuous semi-supervised training between the two, in which the CSS model is further trained by using real data based on the teacher-student learning framework.
We apply this scheme to an array-geometry-agnostic CSS model, which can use the multi-channel data collected from any microphone array.
Large-scale meeting transcription experiments are carried out on both Microsoft internal meeting data and the AMI meeting corpus.
The steady improvement by each training stage has been observed, showing the effect of the proposed method that enables leveraging real conversational data for CSS model training.

\end{abstract}
\noindent\textbf{Index Terms}: Continuous speech separation, array-geometry-agnostic, semi-supervised training, real unsupervised data

\section{Introduction}

Transcribing human-to-human conversations has been a long-standing problem in the speech community. It requires recognizing naturally overlapping utterances of an unknown number of speakers and encompasses various speech processing challenges. Recently, several relevant benchmarking tasks~\cite{watanabe2020chime, chen2020continuous, fu2021aishell, yu2021m2met} have been released especially for meeting scenarios, creating a renewed interest in this interdisciplinary problem~\cite{yoshioka2018recognizing, raj2021integration}.

Continuous speech separation (CSS)~\cite{chen2020continuous, yoshioka2018recognizing, raj2021integration} was proposed to decouple front-end processing from downstream tasks such as automatic speech recognition (ASR)~\cite{li2021recent} and speaker diarization~\cite{park2022review, xiao2021microsoft} while enabling streaming processing. CSS splits a continuously captured long-form conversational audio signal into multiple overlap-free signals, allowing conventional ASR and diarization methods to be used subsequently. This is typically realized with window-based processing, where a speech separation neural network model is used to process each windowed signal. CSS can be performed with either single microphone~\cite{wu2021investigation, chen2021continuous} or a microphone array~\cite{yoshioka2018recognizing, yoshioka2021Vararray} with the latter being more effective in many cases. 


Training multi-channel CSS models requires careful consideration of the training data. 
The common practice is to simulate mixed signals from clean speech samples because the ground-truth clean signals are needed for supervised learning. 
However, this inevitably causes a mismatch between the training and deployment environments. 
An ASR-loss-based training approach~\cite{yoshioka2021Vararray, chang2019mimo, subramanian2019investigation, wang2021exploring} attempts to address this issue by using ground-truth transcriptions and an ASR model to define the loss function to minimize. Nonetheless, acquiring accurate transcriptions during speaker-overlapping periods is challenging even for humans, making this approach difficult to scale. 
In addition, multi-channel recordings may be obtained from different devices with different microphone array geometries, creating another challenge to utilize the real conversational data.

Due to the data acquisition challenge mentioned above, it is desirable to
leverage real conversational data without any supervision signals.
For single-channel speech enhancement and separation tasks, various techniques to leverage unsupervised data have been investigated, 
including cycle-consistency-loss-based training~\cite{wang2020SSLSE,xiang2020CCGANSE}, 
semi-supervised training~\cite{xia2021realdataSE,tzinis2022remixit}, WavLM~\cite{chen2021wavlm}, and mixture invariant training~\cite{wisdom2020unsupervised}.
However, 
no work has been done to leverage real conversational unsupervised data for the {\it multi-channel speech separation} modeling, where the model must learn the nonlinear correlation between the input channels while solving the permutation problem of the multiple output signals.
In addition, the existing studies used artificial pre-segmented speech mixtures for evaluation rather than real conversational data. Hence, whether these approaches are useful for real applications remains to be seen.

In this paper, we carry out a comprehensive study of approaches that leverage the real-recording training data for multi-channel CSS. To this end, we propose a three-stage training framework: (i) pre-training
using simulated data, (ii) continuous semi-supervised training using unlabeled real data, and (iii) ASR-loss-based fine-tuning using transcribed real data. 
This framework allows us to fully leverage both a large amount of unlabeled real data and a small amount of high-quality transcribed real data.
We use an array-geometry-agnostic multi-channel separation model \cite{yoshioka2021Vararray} to utilize multi-channel recordings from different microphone arrays. 
Unlike existing studies, we use real meeting recordings, i.e.,  Microsoft internal meetings \cite{yoshioka2019advances} and the AMI corpus \cite{Carletta06}, and attempt to obtain insights that are relevant to real application scenarios. 

\begin{figure*}[t]
  \centering
  \includegraphics[width=1.0\linewidth]{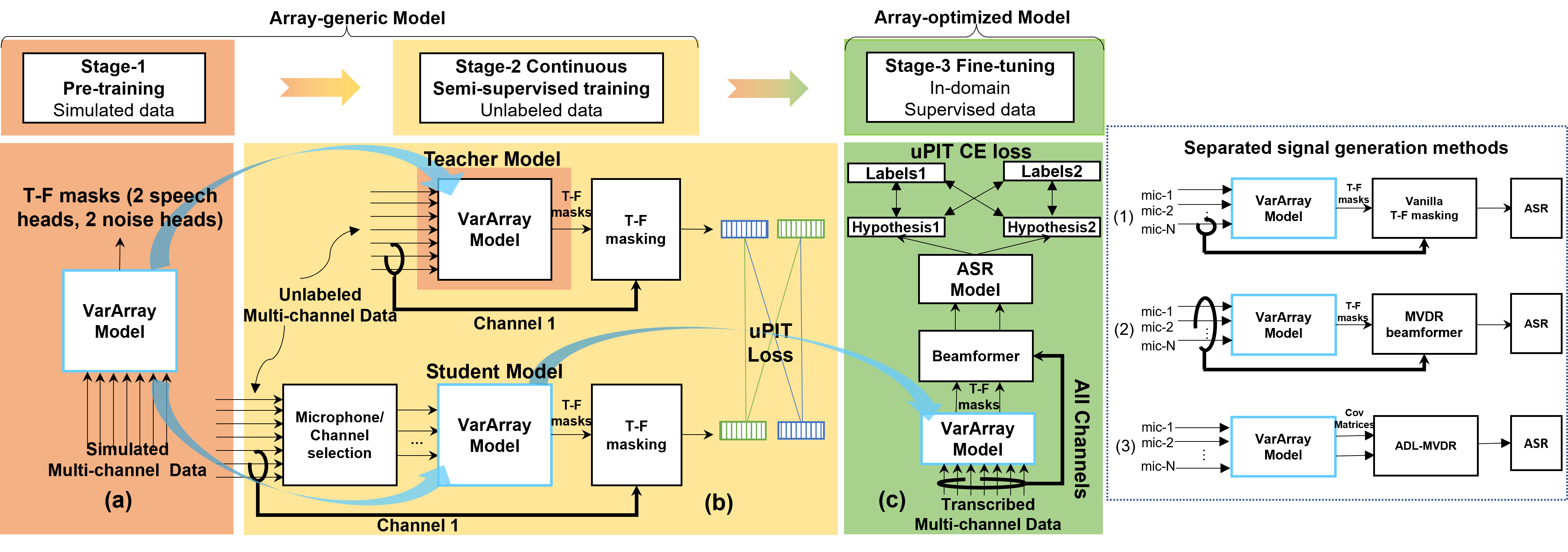}
  \vspace{-7mm}
      \caption{An overview of the system for multi-channel CSS model training and evaluations.}
  \label{fig:system}
  \vspace{-5mm}
\end{figure*}

\section{Background work}

\subsection{VarArray model}
\label{sec:Vararray}
\vspace{-.3em}
Since real multi-channel recordings may be obtained from different devices with different microphone array geometries, it is desirable for a speech separation model to be able to deal with any array geometries without modification or retraining. Therefore, we make use of a recently proposed array-geometry-agnostic model, called VarArray~\cite{yoshioka2021Vararray}, in our investigation. 

VarArray interleaves conformer blocks and cross-channel layers to model both temporal and spatial information of the multi-channel signals. It intakes a multi-channel short-time Fourier transform (STFT) sequence and outputs the time-frequency (T-F) masks for two speech sources and two noise sources (i.e., stationary and transient noises). This is performed in such a way that the outputs stay the same for any permutations of the microphones. This is realized by using transform-average-concatenate (TAC) for the cross-channel layers. The model was shown to significantly outperform a conventional array-geometry-agnostic separation model, which fused the channel outputs only at the last layer~\cite{heymann2016neural}. The CSS framework can be used to process long-form audio. That is, the model is applied for a short-windowed input by sliding the window, followed by a window-stitching procedure~\cite{yoshioka2019advances} to concatenate the short separated signals. Refer to \cite{yoshioka2021Vararray} for more details.

\subsection{Separated signal generation methods}
\label{sec:evalsystem}
\vspace{-.3em}

The T-F masks generated by the speech separation model can be used in different ways to produce the separated speech signals. Multiple methods were proposed in the past, and the effect of the real-data-based training may vary depending on the separated signal generation method. The following three methods are considered in this work. 

The first method is vanilla T-F masking, which directly multiplies the first-channel STFT of the input with the T-F masks (Fig.\ref{fig:system}-(1)). The obtained STFTs are converted to time-domain signals and fed to an ASR system. While this method is dominant for single-channel processing, it often causes speech distortion, limiting the benefit of speech separation. The second approach is minimum variance distortion-less response (MVDR) beamforming. The MVDR filters are estimated from the spatial co-variance matrices calculated with the observed STFTs and the T-F masks (Fig.\ref{fig:system}-(2))~\cite{heymann2016neural, yoshioka2018multi}. As the third method, we also consider a neural network-based beamforming method called all deep learning (ADL)-MVDR (Fig.\ref{fig:system}-(3))~\cite{zhang2021all}. It should be noted that, although we do not investigate the aspect of processing latency, the first and third methods can be performed on a frame-by-frame basis, while MVDR requires batch processing within each window of the CSS.

\section{Three-stage CSS training}
This section introduces the three-stage training scheme that we propose and use for our investigation. 
This training scheme is designed to fully leverage heterogeneous real data, including large-scale data without reference transcriptions nor clean signals, as well as a small amount of transcribed data.

\subsection{Stage-1: Pre-training with simulated data}
\vspace{-.3em}

In stage-1, which we call pre-training, a VarArray speech separation model is trained based on multi-channel simulation data.
The multi-channel signals are simulated from clean speech signals, noise samples, and room impulse responses so that each training sample contains up to two speakers. 
The model is trained based on the utterance-level permutation-invariant training (uPIT) method~\cite{Kolbaek2017UPIT} and as follows,
\begin{align}
\label{loss1}
\mathcal{L}_{\textrm{stage-1}} = \mathcal{L}_{\textrm{uPIT}} + \sum\limits_{q\in Q}w_q \|M_q \odot|Y|- |N|\ \|,  \\
\mathcal{L}_{\textrm{uPIT}} = \min_{\phi \in P}\sum\limits_{(i, j)\in P }\|M_i \odot|Y| - |X_j|\ \|,
\end{align}
where $(i, j)\in P$ contains all the possible permutations of the two speech sources and $Q$ represents the two types of noises. $M$ and $|Y|$ are T-F masks estimated from VarArray and the magnitude of one observed channel, respectively. $|X|$ and $|N|$ denote the magnitude of the reference speech and noise signals. Operator $\odot$ denotes the elementary multiplication for each T-F bin. $\|\cdot\|$ is the L2 distance and $w_q=0.1$ is the weight for the noise losses. 
This stage is the same as the original VarArray model training, and thus we refer the readers to~\cite{yoshioka2021Vararray}.

\subsection{Stage-2: Continuous semi-supervised training}
\vspace{-.3em}

In stage-2 (Fig.\ref{fig:system}-(b)), the VarArray speech separation model is updated, starting from the stage-1 model, with semi-supervised training~\cite{chapelle2006semi} based on the teacher-student learning framework. To do so, in addition to the stage-1 model (i.e., the student model), we also generate a teacher model based on the simulated training data, where the teacher model can be the same as the student model or a bigger and more accurate one. Given the real-recording training samples, the teacher model generates T-F masks. Then, the student model is updated by using these masks as the learning references. The simulated data may also be used together during the stage-2 training. 

Recall that VarArray has four outputs, with the first two corresponding to the two speech sources and the rest to 
the stationary and transient noise signals. Only the first two outputs have two-way permutation ambiguity, while the order of the two noise outputs is fixed.
Hence, we define the stage-2 loss function $\mathcal{L}_{\textrm{stage-2}}$ as a weighted sum of the uPIT-style loss for the speech sources and the L2 losses for the stationary and transient noise signals. That is,  
\vspace{-.3em}
\begin{align}
\label{loss}
\mathcal{L}_{\textrm{stage-2}} = \mathcal{L}_{\textrm{uPIT}}^{\textrm{T-S}} + \sum\limits_{q\in Q}w_q \|M_q^{\textrm{tea}} \odot|Y|-M_q^{\textrm{stu}}\odot|Y|\ \|,  \\
\mathcal{L}_{\textrm{uPIT}}^{\textrm{T-S}} = \min_{\phi \in P}\sum\limits_{(i, j)\in P }\|M_i^{\textrm{tea}} \odot|Y|-M_j^{\textrm{stu}}\odot|Y|\ \|,
\end{align}
where $M^{\textrm{tea/stu}}$ is the T-F masks estimated from teacher/student.

The long-form audio signals obtained from real conversation recordings, which can be longer than an hour, must be segmented to be used for the training. We examine two segmentation schemes. One method, called fixed-window segmentation (FWS), simply splits the multi-channel long-form audio signals into 4-second pieces without overlaps. With FWS, some segments can occasionally contain more than two speakers, which may confuse the model. The other method, which we refer to as conversational transcription system-based segmentation (CTS), attempts to control the audio portions to be used for the training based on their characteristics. Specifically, the multi-channel conversational transcription system of \cite{yoshioka2019advances} is applied to the signals to obtain word-level speaker diarization results. Then, we sweep the long-form signal from the beginning and cut it (1) when the third speaker is detected, (ii) when the length of the current segment exceeds a threshold (20 seconds in our experiments), or (iii) when there is a silence period longer than a threshold (2.5 seconds). This also allows us to sample two-speaker segments more frequently during training to promote speech separation learning.

In addition, to maintain the model's ability to deal with various microphone arrays and acoustic conditions, a subset of the input channels is randomly chosen for each training sample that is fed to the student model. On the other hand, the teacher model uses all the input signals to generate T-F masks that are as accurate as possible. As a result, the stage-2 trained model is expected to be adapted to real data distributions while largely retaining the array-geometry-agnostic property.

\subsection{Stage-3: ASR-loss-based Fine-tuning}
\vspace{-.3em}
Similar to \cite{yoshioka2021Vararray}, the VarArray model can be fine-tuned further with an ASR-based loss function, where the VarArray-based mask estimation model, MVDR beamformer, gain adjustment, feature transformation, global mean-variance normalization, and ASR model are combined within a single network (Fig.\ref{fig:system}--(c)). 
We use an attention-based encoder-decoder-based ASR model of \cite{wang2021exploring}, which estimates text sequence $C = (c_1, c_2, \cdots)$ from the separated time-domain signal that is generated by the CSS model and the MVDR beamforming. 
In this fine-tuning stage, only parameters of the VarArray model are updated while freezing the other modules. The VarArray model is initialized based on the stage-1 or stage-2 one. All connected modules are differentiable and uPIT-style cross-entropy loss between two predicted hypotheses $H_i=\{h_1,...,h_N, h_{N+1}=\langle eos\rangle\}$ ($i=1$ or $2$, $N$ is the length of $H$ and $\langle eos\rangle$ denotes the end of sequence) and reference transcriptions $C_j$ ($j=1$ or $2$) is back-propagated all the way down to the VarArray.

All channels are used for VarArray fine-tuning, so the model will be forced to learn the geometry information from the training data of a specific array, such that the model is adapted from an array-generic one to a geometry-optimized one.

\section{Experiments}

\subsection{Experiment settings}
\vspace{-.3em}

Details of our training data in each training stage are shown in Table \ref{tab:data}.
Note that we did not use reference segmentation information in the 2nd-stage, although some corpus contains time-aligned transcriptions. Three stage-1 pre-trained VarArray models were used as either teachers or students.
Based on the number of parameters of the models (See Table.\ref{tab:1}), we referred to them as Large (L), Small (S), or eXtra-Small (XS). 
These three models shared the same model architecture used in \cite{yoshioka2021Vararray} except 
that the number of modules (conformer blocks and TAC layers) and layer nodes were differently configured. 
Specifically, 5 conformer blocks in total were used for L and S, and 3 were used for XS.
The conformer block comprised 5 conformer layers with 10, 4, and 3 attention heads, 33 convolution kernels, 512, 64, and 48 dimensions for L, S, and XS models, respectively.  
For the stage-2 training, the learning rate was set to $1e^{-4}$ and exponentially decayed at a rate of $0.99998$. Adam optimizer was used with a weight decay of $1e^{-5}$, while the stage-1 and stage-3 setups were the same as our previous work \cite{yoshioka2021Vararray}.
CTS was used for audio segmentation at stage-2 unless otherwise stated.

In evaluation,
we used Microsoft internal meeting collections (MS-eval) \cite{yoshioka2019advances}, consisting of 60 sessions of real meeting recordings (150K words in the reference transcriptions), and AMI-MDM development and evaluation sets (AMI-dev/eval).
We used 
the window length and window shift rate of 1.6 seconds and 0.4 seconds, respectively, for CSS
in all evaluation systems. 
Microsoft's internal ASR system was used to generate transcriptions from two separation outputs.
We evaluated each system based on the word error rate (WER).
\vspace{-.3em}
\begin{table}[h]
  \begin{center}
   	\caption{Training data used in the experiments.
   	 }
    \label{tab:data}
    \vspace{-3mm}
    \resizebox{0.48\textwidth}{!}{
    {\scriptsize 
	\begin{tabular}{c|c|c|c|c}
	  \toprule
	  {\bf Stage} & {\bf Name} & {\bf Size (hr)} & {\bf Channel} & {\bf Short Description}  \\ \hline
      1 & SIM~\cite{yoshioka2021Vararray} & 3,000 & 7 or 8 & Simulated by WSJ, LibriSpeech \\ \hline
      \multirow{6}{*}{2} & MS & 875 & 7 & Meeting recordings at Microsoft \\
       & MS$^{\rm sub}$ & 450 & 7 &  Subset of MS\\
       & AMI~\cite{Carletta06} & 80 & 8 &  English Meeting recordings \\
       & AISHELL-4~\cite{fu2021aishell} & 107 & 8 & Chinese Meeting recordings \\
       & AliMeeting~\cite{Yu2022M2MeT} & 104 & 8 &  Chinese Meeting recordings \\
       & SIM$^{\rm sub}$~\cite{yoshioka2021Vararray} & 219 & 7 & Subset of SIM\\ \hline 
      3 & AMI$^{\rm sub}$~\cite{Carletta06} & 8.52 & 8 &  Subset of up-to-two concurrent speakers\\
      \bottomrule
	\end{tabular}
}
}
  \end{center}
 \vspace{-6mm}
\end{table}


\subsection{Effect of stage-2 continuous semi-supervised training}
\vspace{-.3em}

\subsubsection{In-domain experiments}
\vspace{-.3em}
We first examined the stage-2 training based on the matched condition for training (MS) and testing (MS-eval).
Table \ref{tab:1} shows the WERs using different teacher and student models, evaluated by the vanilla T-F masking system (Fig.\ref{fig:system}-(1)).
As expected, with only stage-1 pre-training, the larger models 
yielded lower WERs at the expense of increased computational cost, indicated by the real-time factor (RTF).
Note that L and S models were applied to all microphones during evaluation while XS was applied to three of them.
Comparing the second and third rows shows that the continuous semi-supervised training largely improved the WERs for both non-overlap (Non-Ovlp) and overlap (Ovlp) regions, even when the teacher and student were the same models.
This indicates that continuous semi-supervised training forced the CSS model to learn the real data characteristics.
Using a larger teacher model in semi-supervised training brought more gains (third and fourth rows) for the student model, 
which is achieved by the additional effect of knowledge distillation.

\begin{table}[t]
\centering
\caption{Impact of continuous semi-supervised training with different model sizes. Vanilla T-F masking was used.}
\vspace{-2mm}
\label{tab:1}
\resizebox{0.48\textwidth}{!}{
\begin{tabular}{c|c|c|c|ccc}
\toprule
 \multirow{2}{*}{\bf Teacher} & \multirow{2}{*}{\bf Student} & \multirow{2}{*}{\shortstack[l]{\bf Num. of\\{\bf param.}}} & \multirow{2}{*}{\shortstack[c]{\bf RTF$^\dagger$}} & \multicolumn{3}{c}{\bf MS-eval WER}  \\ 
 & & & & {\bf Ovlp} & {\bf Non-Ovlp} & {\bf Overall}  \\ \hline
 - & L  & 54.56M & 3.0    & 31.5 & 15.5 & 19.4     \\
 - & S  & 1.04M & 0.24 & 33.3 & 16.7 & 20.8      \\
 S & S  & 1.04M &  0.24 & 32.1 & 16.1 & 20.0     \\
 L & S  & 1.04M &  0.24 & {\bf 31.6} & {\bf 15.8} & {\bf 19.7}     \\
 \hline
 - & XS   & 0.34M  & 0.045 & 35.1 & 17.5 & 21.8    \\
 L & XS   & 0.34M  & 0.045 & {\bf 33.5} & {\bf 16.3} & {\bf 20.5}    \\
\bottomrule
\end{tabular}
}
\begin{tablenotes}
\scriptsize
\item[*] \hspace{1mm}$^\dagger$ The RTF was measured with Intel Xeon machine clocked at 2.1 GHz with single thread setting.
\end{tablenotes}
\end{table}

Table \ref{tab:2} compares the effects of the semi-supervised training on different evaluation systems, where the L and S models were used as the teacher and student, respectively. 
We can see that the continuous semi-supervised training consistently improved the WERs for all the evaluation systems. 
Better WERs were achieved by the MVDR beamformer and ADL-MVDR, although the relative gains for these systems were smaller than for the masking-based systems. 
This can be because the beamformers are less sensitive to mask estimation errors. 

\begin{table}[t]
\centering
\caption{Impact of continuous semi-supervised training on evaluation systems (see section \ref{sec:evalsystem}).}
\vspace{-2mm}
\label{tab:2}
\resizebox{0.45\textwidth}{!}{
\begin{tabular}{c|c|ccc}
\toprule
 \multirow{2}{*}{\bf Block-online evaluation system} & \multirow{2}{*}{\bf Stage-2} & \multicolumn{3}{c}{\bf MS-eval WER}   \\ 
 &  & {\bf Ovlp} & {\bf Non-Ovlp} & {\bf Overall} \\ \hline
 \multirow{2}{*}{Vanilla T-F Masking}       
 &  -             & 33.3 & 16.7 & 20.8    \\
 &  \checkmark    & {\bf 31.6} & {\bf 15.8} & {\bf 19.7}    \\
 \hline
 \multirow{2}{*}{MVDR beamformer}     
 &  -             & 24.7 & 12.7  & 15.6    \\
 &  \checkmark    & {\bf 24.3} & {\bf 12.6}  & {\bf 15.5}  \\
 \hline
 \multirow{2}{*}{ADL-MVDR~\cite{zhang2021all}} 
  & -             & 25.8 & 13.2 & 16.3    \\
  & \checkmark    & {\bf 25.3} & {\bf 12.7} & {\bf 15.7}    \\
\bottomrule
\end{tabular}
}
\vspace{-5mm}
\end{table}

\subsubsection{Cross-domain experiments}
\vspace{-.3em}

Table \ref{tab:3} shows the cross-validation results between the MS and AMI.
L teacher, S student, and vanilla T-F masking system were used in these experiments.
Firstly, by comparing the 1st row with 2nd and 3rd rows, we observed that stage-2 training always improved the WER even when the array configurations for the training data and testing data were different, supporting the importance of real training data.
Secondly, by comparing the 4-th and 5-th rows, we observed a significant improvement from stage-1 training,
which indicates the necessity of pre-training using simulated data.
Thirdly, as expected, the best WERs of MS and AMI test sets were achieved by using the matched training sets as in the 2nd and 3rd rows (6.3\% and 5.2\% relative WER reduction for MS and AMI, respectively). 
It is noteworthy that the WER difference between the array-mismatched and array-matched conditions was not significant, indicating that the stage-2-trained models largely maintained the array-geometry-agnostic property.
Finally, from 3rd, 5th, and 6th rows, it can be inferred that using more unlabeled data resulted in more gains in WERs while using CTS could improve the model performance for overlap regions (32.3\% vs. 31.6\%). 
This was because the CTS filtered out most of the silence part in the long-form audios and tended to form more proportions of overlapped segments for training. 
\begin{table}[t]
\centering
\caption{Impact of training data on \%WER of MS and AMI sets. Vanilla T-F masking was used.}
\vspace{-2mm}
\label{tab:3}
\resizebox{0.48\textwidth}{!}{
\begin{tabular}{c|c|c|c|ccc|cc}
\toprule
{\bf Stage-}&{\bf Stage-}  & \multirow{2}{*}{\bf Segment} & {\bf Stage-2}  & \multicolumn{3}{c|}{\bf MS-eval WER} & \multicolumn{2}{c}{\bf AMI-dev/eval WER}  \\ 
 {\bf 1} & {\bf 2} & & {\bf hours}& {\bf Ovlp} & {\bf Non-Ovlp} & {\bf Overall} & {\bf Non-Ovlp} & {\bf Overall}  \\ 
\hline
SIM & - & - & 0 & 33.3 & 16.7 & 20.8  & 20.2 / 21.3  & 26.7 / 29.0 \\
SIM & AMI        &    FWS            & 80       & 32.6 & {\bf 15.4} & 19.6 & {\bf 18.4} / {\bf 20.0} & {\bf 24.9} / {\bf 27.5}   \\
SIM & MS          &    FWS           &  875     & 32.3 & {\bf 15.4} & {\bf 19.5} & 19.2 / 20.1 & 25.4 / 27.9  \\
\hline
 - & MS$^{\rm sub}$ &     CTS  &  234     & 36.9 & 19.6 & 23.9 & 21.6 / 24.1        & 29.5 / 32.6            \\
SIM & MS$^{\rm sub}$   &    CTS &  234     & 32.3 & 16.0 & 20.0 & 19.1 / 20.6 & 25.8 / 28.0  \\
SIM & MS          &    CTS  &  647     & {\bf 31.6} & 15.8 & 19.7 & 19.3 / 20.5 & 25.6 / 28.0  \\
\bottomrule
\end{tabular}
}
\vspace{-2mm}
\end{table}

We further investigated the impact of the training data (based on FWS) on stage-2 training. Specifically, we performed the stage-2 training on data sets obtained from different corpora and evaluated the resultant models on the AMI dev and eval sets. 
To remove outlier data samples while ensuring a reasonable data quantity for each corpus, we selected the training data by discarding the samples with DNSMOS scores smaller than 2.5, except for the MS set, for which we used the threshold of 3.0. DNSMOS is a neural network that predicts the speech quality without requiring the ground-truth speech labels, which was proposed in \cite{reddy2021DNSMOS}. 

Table \ref{tab:languageEff} shows the result. By comparing the first three rows of the stage-2 models, we can observe that the language mismatch between stage-2 training and evaluation had no negative impact on the performance. However, the model trained on the English-based MS set slightly underperformed the models trained on either AMI or AIShell + AliMeetings. This indicates that the VarArray-based speech separation model seems to largely rely on spatial acoustics rather than the phonetic aspects. Combining all the training sets, including the simulated data, achieved the best accuracy on average. Including the simulated samples had a positive impact only on the dev set.

\begin{table}[h]
\caption{Additional stage-2 training data investigation on AMI. Vanilla T-F masking was used.}
\vspace{-2mm}
\label{tab:languageEff}
\resizebox{0.48\textwidth}{!}{
\begin{tabular}{cccc|lc|c|c}
\toprule
 \multicolumn{4}{c|}{\textbf{Stage-2 training data}} &
  \multicolumn{2}{c|}{\textbf{Lang}} &
  \textbf{Stage-2}&
  \multicolumn{1}{c}{\textbf{AMI WER}} \\
\multicolumn{1}{l}{\textbf{AMI}} &
  \multicolumn{1}{l}{\textbf{AISH+Ali}} &
  \multicolumn{1}{l}{\textbf{MS}} &
  \textbf{SIM$^{\rm sub}$} &
  \multicolumn{1}{l}{\textbf{EN}} &
  \textbf{CH} & \textbf{hours}
   &
  \textbf{dev/eval} \\ \hline
\multicolumn{4}{c|}{Baseline-Stage-1} & \checkmark   &   & 0 & 26.7 / 29.0            \\  \hline
\checkmark &            &            &            & \checkmark &            & 44   & 25.2 / 28.0                                 \\
           & \checkmark &            &            &            & \checkmark & 148  & 25.2 / \textbf{27.7}                         \\
           &            & \checkmark &            & \checkmark &            & 359   & 25.6 / 28.0                         \\
\checkmark & \checkmark & \checkmark &            & \checkmark & \checkmark & 551  & 25.2 / 27.8      \\ 
\checkmark & \checkmark & \checkmark & \checkmark & \checkmark & \checkmark & 770  & \textbf{24.9} / 27.8      \\ 
\bottomrule
\end{tabular}
}
\vspace{-5mm}
\end{table}

\subsection{Effect of stage-3 ASR-loss-based fine-tuning}
\vspace{-0.3em}
The result with and without the stage-3 ASR-loss-based fine-tuning is shown in Table \ref{tab:5}. Refer \cite{wang2021exploring, eskimez2021human} for the details of the ASR model used in this experiment. 
We fine-tuned the stage-1 VarArray model trained by SIM, or stage-2 VarArray model pre-trained by SIM and continuously trained by AMI without the labels.
From the result, the stage-3 fine-tuning always brought us improvements,
and the best result was obtained for the model trained from the stage-2 model.
This result supports the importance of all training stages we proposed in the paper.


\vspace{-0.3em}
\begin{table}[h]
\centering
\caption{Impact of stage-3 fine-tuning on \%WER. MVDR beamformer was used.}
\vspace{-2mm}
\label{tab:5}
\resizebox{0.48\textwidth}{!}{
\begin{tabular}{c|c|cc|cc}
\toprule
\multirow{2}{*}{\bf Pre-Train} & \multirow{2}{*}{\bf Stage-3} & \multicolumn{2}{c|}{\bf AMI-dev WER}  & \multicolumn{2}{c}{\bf AMI-eval WER} \\ 
 &  & {\bf Non-Ovlp} & {\bf Overall} & {\bf Non-Ovlp} & {\bf Overall} \\ \hline
Stage-1 &                    &  15.0 & 17.6 & 16.6 & 20.4     \\
Stage-1 &         \checkmark & {\bf 14.6} & {\bf 17.4} & {\bf 16.3} & {\bf 20.3} \\ 
Stage-1+Stage-2 &            &  14.5 & 17.2 & 16.5 & 20.2      \\
Stage-1+Stage-2 & \checkmark & {\bf 14.2} & {\bf 17.0} & {\bf 16.3} & {\bf 20.1} \\ 
\bottomrule
\end{tabular}
}
\vspace{-5mm}
\end{table}

\section{Conclusions}
In this work, a three-stage training scheme for a multi-channel array-geometry-agnostic CSS model was introduced. 
Specifically, we proposed to train the CSS model in the order of stage-1 supervised training using simulated data, stage-2 continuous semi-supervised training using real unlabeled data, and stage-3 ASR-loss-based fine-tuning based on real transcribed data. 
Through the streaming conversational transcription experiments using a mix of internal and public data, we observed that
(i) each training stage contributed steady WER improvement;
(ii) the stage-2 training could significantly reduce the train-test mismatch issue that occurred in stage-1 and yielded a better CSS model for further in-domain stage-3 fine-tuning and downstream ASR; 
(iii) real unsupervised data, regardless of the array geometry and language, was beneficial not only for the in-domain inference but also for out-of-domain ones.

\newpage
\bibliographystyle{IEEEtran}
\bibliography{myref}

\begin{thebibliography}{10}
\providecommand{\url}[1]{#1}
\csname url@samestyle\endcsname
\providecommand{\newblock}{\relax}
\providecommand{\bibinfo}[2]{#2}
\providecommand{\BIBentrySTDinterwordspacing}{\spaceskip=0pt\relax}
\providecommand{\BIBentryALTinterwordstretchfactor}{4}
\providecommand{\BIBentryALTinterwordspacing}{\spaceskip=\fontdimen2\font plus
\BIBentryALTinterwordstretchfactor\fontdimen3\font minus
  \fontdimen4\font\relax}
\providecommand{\BIBforeignlanguage}[2]{{%
\expandafter\ifx\csname l@#1\endcsname\relax
\typeout{** WARNING: IEEEtran.bst: No hyphenation pattern has been}%
\typeout{** loaded for the language `#1'. Using the pattern for}%
\typeout{** the default language instead.}%
\else
\language=\csname l@#1\endcsname
\fi
#2}}
\providecommand{\BIBdecl}{\relax}
\BIBdecl

\bibitem{watanabe2020chime}
S.~Watanabe, M.~Mandel, J.~Barker, E.~Vincent, A.~Arora, X.~Chang,
  S.~Khudanpur, V.~Manohar, D.~Povey, D.~Raj \emph{et~al.}, ``Chime-6
  challenge: Tackling multispeaker speech recognition for unsegmented
  recordings,'' \emph{arXiv preprint arXiv:2004.09249}, 2020.

\bibitem{chen2020continuous}
Z.~Chen, T.~Yoshioka, L.~Lu, T.~Zhou, Z.~Meng, Y.~Luo, J.~Wu, X.~Xiao, and
  J.~Li, ``Continuous speech separation: Dataset and analysis,'' in \emph{Proc.
  ICASSP}.\hskip 1em plus 0.5em minus 0.4em\relax IEEE, 2020, pp. 7284--7288.

\bibitem{fu2021aishell}
Y.~Fu, L.~Cheng, S.~Lv, Y.~Jv, Y.~Kong, Z.~Chen, Y.~Hu, L.~Xie, J.~Wu, H.~Bu
  \emph{et~al.}, ``Aishell-4: An open source dataset for speech enhancement,
  separation, recognition and speaker diarization in conference scenario,''
  \emph{arXiv preprint arXiv:2104.03603}, 2021.

\bibitem{yu2021m2met}
F.~Yu, S.~Zhang, Y.~Fu, L.~Xie, S.~Zheng, Z.~Du, W.~Huang, P.~Guo, Z.~Yan,
  B.~Ma \emph{et~al.}, ``{M2MeT}: The {ICASSP} 2022 multi-channel multi-party
  meeting transcription challenge,'' \emph{arXiv preprint arXiv:2110.07393},
  2021.

\bibitem{yoshioka2018recognizing}
T.~Yoshioka, H.~Erdogan, Z.~Chen, X.~Xiao, and F.~Alleva, ``Recognizing
  overlapped speech in meetings: A multichannel separation approach using
  neural networks,'' \emph{Proc. Interspeech}, pp. 3038--3042, 2018.

\bibitem{raj2021integration}
D.~Raj, P.~Denisov, Z.~Chen, H.~Erdogan, Z.~Huang, M.~He, S.~Watanabe, J.~Du,
  T.~Yoshioka, Y.~Luo \emph{et~al.}, ``Integration of speech separation,
  diarization, and recognition for multi-speaker meetings: System description,
  comparison, and analysis,'' in \emph{Proc. SLT}.\hskip 1em plus 0.5em minus
  0.4em\relax IEEE, 2021, pp. 897--904.

\bibitem{li2021recent}
J.~Li, ``Recent advances in end-to-end automatic speech recognition,''
  \emph{arXiv preprint arXiv:2111.01690}, 2021.

\bibitem{park2022review}
T.~J. Park, N.~Kanda, D.~Dimitriadis, K.~J. Han, S.~Watanabe, and S.~Narayanan,
  ``A review of speaker diarization: Recent advances with deep learning,''
  \emph{Computer Speech \& Language}, vol.~72, p. 101317, 2022.

\bibitem{xiao2021microsoft}
X.~Xiao, N.~Kanda, Z.~Chen, T.~Zhou, T.~Yoshioka, S.~Chen, Y.~Zhao, G.~Liu,
  Y.~Wu, J.~Wu \emph{et~al.}, ``Microsoft speaker diarization system for the
  {VoxCeleb} speaker recognition challenge 2020,'' in \emph{Proc.
  ICASSP}.\hskip 1em plus 0.5em minus 0.4em\relax IEEE, 2021, pp. 5824--5828.

\bibitem{wu2021investigation}
J.~Wu, Z.~Chen, S.~Chen, Y.~Wu, T.~Yoshioka, N.~Kanda, S.~Liu, and J.~Li,
  ``Investigation of practical aspects of single channel speech separation for
  {ASR},'' \emph{arXiv preprint arXiv:2107.01922}, 2021.

\bibitem{chen2021continuous}
S.~Chen, Y.~Wu, Z.~Chen, J.~Wu, J.~Li, T.~Yoshioka, C.~Wang, S.~Liu, and
  M.~Zhou, ``Continuous speech separation with conformer,'' in \emph{Proc.
  ICASSP}.\hskip 1em plus 0.5em minus 0.4em\relax IEEE, 2021, pp. 5749--5753.

\bibitem{yoshioka2021Vararray}
T.~Yoshioka, X.~Wang, D.~Wang, M.~Tang, Z.~Zhu, Z.~Chen, and N.~Kanda,
  ``{VarArray}: Array-geometry-agnostic continuous speech separation,''
  \emph{arXiv preprint arXiv:2110.05745}, 2021.

\bibitem{chang2019mimo}
X.~Chang, W.~Zhang, Y.~Qian, J.~Le~Roux, and S.~Watanabe, ``{MIMO-SPEECH}:
  End-to-end multi-channel multi-speaker speech recognition,'' in \emph{Proc.
  ASRU}.\hskip 1em plus 0.5em minus 0.4em\relax IEEE, 2019, pp. 237--244.

\bibitem{subramanian2019investigation}
A.~S. Subramanian, X.~Wang, S.~Watanabe, T.~Taniguchi, D.~Tran, and Y.~Fujita,
  ``An investigation of end-to-end multichannel speech recognition for
  reverberant and mismatch conditions,'' \emph{arXiv preprint
  arXiv:1904.09049}, 2019.

\bibitem{wang2021exploring}
X.~Wang, N.~Kanda, Y.~Gaur, Z.~Chen, Z.~Meng, and T.~Yoshioka, ``Exploring
  end-to-end multi-channel {ASR} with bias information for meeting
  transcription,'' in \emph{Proc. SLT}.\hskip 1em plus 0.5em minus 0.4em\relax
  IEEE, 2021, pp. 833--840.

\bibitem{wang2020SSLSE}
Y.-C. Wang, S.~Venkataramani, and P.~Smaragdis, ``Self-supervised learning for
  speech enhancement,'' \emph{ICML}, 2020.

\bibitem{xiang2020CCGANSE}
Y.~Xiang and C.~Bao, ``A parallel-data-free speech enhancement method using
  multi-objective learning cycle-consistent generative adversarial network,''
  \emph{IEEE/ACM TRANSACTIONS ON AUDIO, SPEECH, AND LANGUAGE PROCESSING},
  vol.~28, pp. 1826--1838, 2020.

\bibitem{xia2021realdataSE}
Y.~Xia, B.~Xu, and A.~Kumar, ``Incorporating real-world noisy speech in
  neural-network-based speech enhancement systems,'' \emph{arXiv preprint
  arXiv:2109.05172v2}, 2021.

\bibitem{tzinis2022remixit}
E.~Tzinis, Y.~Adi, V.~K. Ithapu, B.~Xu, P.~Smaragdis, and A.~Kumar, ``Remixit:
  Continual self-training of speech enhancement models via bootstrapped
  remixing,'' \emph{arXiv preprint arXiv:2202.08862}, 2022.

\bibitem{chen2021wavlm}
S.~Chen, C.~Wang, Z.~Chen, Y.~Wu, S.~Liu, Z.~Chen, J.~Li, N.~Kanda,
  T.~Yoshioka, X.~Xiao \emph{et~al.}, ``{WavLM}: Large-scale self-supervised
  pre-training for full stack speech processing,'' \emph{arXiv preprint
  arXiv:2110.13900}, 2021.

\bibitem{wisdom2020unsupervised}
S.~Wisdom, E.~Tzinis, H.~Erdogan, R.~Weiss, K.~Wilson, and J.~Hershey,
  ``Unsupervised sound separation using mixture invariant training,''
  \emph{Advances in Neural Information Processing Systems}, vol.~33, pp.
  3846--3857, 2020.

\bibitem{yoshioka2019advances}
T.~Yoshioka, I.~Abramovski, C.~Aksoylar, Z.~Chen, M.~David, D.~Dimitriadis,
  Y.~Gong, I.~Gurvich, X.~Huang, Y.~Huang \emph{et~al.}, ``Advances in online
  audio-visual meeting transcription,'' in \emph{Proc. ASRU}.\hskip 1em plus
  0.5em minus 0.4em\relax IEEE, 2019, pp. 276--283.

\bibitem{Carletta06}
J.~Carletta, S.~Ashby, S.~Bourban, M.~Flynn, M.~Guillemot, T.~Hain, J.~Kadlec,
  V.~Karaiskos, W.~Kraaij, M.~Kronenthal, G.~Lathoud, M.~Lincoln, A.~Lisowska,
  I.~McCowan, W.~P. andD. Reidsma, and P.~Wellner, ``The {AMI} meeting corpus:
  a pre-announcement,'' in \emph{Proc. Int. Worksh. Machine Learning for
  Multimodal Interaction}, 2006, pp. 28--39.

\bibitem{heymann2016neural}
J.~Heymann, L.~Drude, and R.~Haeb-Umbach, ``Neural network based spectral mask
  estimation for acoustic beamforming,'' in \emph{2016 IEEE International
  Conference on Acoustics, Speech and Signal Processing (ICASSP)}.\hskip 1em
  plus 0.5em minus 0.4em\relax IEEE, 2016, pp. 196--200.

\bibitem{yoshioka2018multi}
T.~Yoshioka, H.~Erdogan, Z.~Chen, and F.~Alleva, ``Multi-microphone neural
  speech separation for far-field multi-talker speech recognition,'' in
  \emph{Proc. ICASSP}.\hskip 1em plus 0.5em minus 0.4em\relax IEEE, 2018, pp.
  5739--5743.

\bibitem{zhang2021all}
Z.~Zhang, T.~Yoshioka, N.~Kanda, Z.~Chen, X.~Wang, D.~Wang, and S.~E. Eskimez,
  ``All-neural beamformer for continuous speech separation,'' \emph{arXiv
  preprint arXiv:2110.06428}, 2021.

\bibitem{Kolbaek2017UPIT}
M.~Kolb{\ae}k, D.~Yu, Z.-H. Tan, and J.~Jensen, ``Multitalker speech separation
  with utterance-level permutation invariant training of deep recurrent neural
  networks,'' \emph{IEEE/ACM Transactions on Audio, Speech, and Language
  Processing}, vol.~25, no.~10, pp. 1901--1913, 2017.

\bibitem{chapelle2006semi}
O.~Chapelle, B.~Scholkopf, and A.~Zien, \emph{Semi-supervised learning}.\hskip
  1em plus 0.5em minus 0.4em\relax MIT press, 2006.

\bibitem{Yu2022M2MeT}
F.~Yu, S.~Zhang, Y.~Fu, L.~Xie, S.~Zheng, Z.~Du, W.~Huang, P.~Guo, Z.~Yan,
  B.~Ma, X.~Xu, and H.~Bu, ``M2{M}e{T}: The {ICASSP} 2022 multi-channel
  multi-party meeting transcription challenge,'' in \emph{Proc. ICASSP}.\hskip
  1em plus 0.5em minus 0.4em\relax IEEE, 2022.

\bibitem{reddy2021DNSMOS}
C.~K.~A. Reddy, V.~Gopal, and R.~Cluter, ``{DNSMOS P.835}: A non-intrusive
  perceptual objective speech quality metric to evaluate noise suppressors,''
  \emph{arXiv preprint arXiv:2110.01763v2}, 2021.

\bibitem{eskimez2021human}
S.~E. Eskimez, X.~Wang, M.~Tang, H.~Yang, Z.~Zhu, Z.~Chen, H.~Wang, and
  T.~Yoshioka, ``{Human Listening and Live Captioning: Multi-Task Training for
  Speech Enhancement},'' in \emph{Proc. Interspeech 2021}, 2021, pp.
  2686--2690.

\end{thebibliography}

\end{document}